\documentclass[final,5p,twocolumn,number,sort&compress]{elsarticle}
\usepackage{graphicx}
\usepackage{amsmath,amssymb}
\usepackage[utf8]{inputenc}
\usepackage{psfrag}
\usepackage{bm}
\newcommand{\chic}{\chi_{\scriptstyle \rm c}}

\begin{document}
\begin{frontmatter}
\title{Intransitivity and coexistence in four species cyclic games}

\author[ufrgs]{Alessandra F. L\"utz}
\ead{sandiflutz@gmail.com}
\author[cab]{Sebastián Risau-Gusman}
\ead{srisau@cab.cnea.gov.ar}
\author[ufrgs,lpthe]{Jeferson J. Arenzon\corref{cor1}}
\ead{arenzon@if.ufrgs.br}

\cortext[cor1]{Corresponding author, +55 51 33086478}
\address[ufrgs]{Instituto de F\'\i sica, Universidade Federal do 
Rio Grande do Sul, CP 15051, 91501-970 Porto Alegre RS, Brazil} 
\address[cab]{Consejo Nacional de Investigaciones Científicas y Técnicas, Centro Atómico Bariloche, 8400 San Carlos de 
Bariloche, Río Negro, Argentina}
\address[lpthe]{LPTHE, Universit\'e Pierre et Marie Curie-Paris VI, 4 Place Jussieu, 75252 Paris Cedex 05, France}

\date{\today}

\begin{abstract}
Intransitivity is a property of connected, oriented graphs
representing species interactions that may drive their coexistence  
even in the presence of competition, the standard example being the 
three species Rock-Paper-Scissors game. We consider here a generalization with four
species, the minimum number of species allowing other
interactions beyond the single loop (one predator, one prey).
We show that, contrary to the mean field prediction, on a square lattice
the model presents a transition, as the parameter setting the
rate at which one species invades another changes, from a coexistence to 
a state in which one species gets extinct. Such a dependence on the
invasion rates shows that the interaction graph structure alone is not
enough to predict the outcome of such models. In addition,
different invasion rates permit to tune the level of transitiveness, indicating
that for the coexistence of all species to persist, 
there must be a minimum amount of intransitivity.
\end{abstract}


\end{frontmatter}

%\maketitle

\section{Introduction}

Cyclic competition~\cite{HoSi98,SzFa07,Frey10} among a population of ${\cal S}$ species (or 
${\cal S}$ different traits within a species) may occur when the trophic network presents 
loops, for which several examples exist: mating lizards~\cite{SiLi96}, competing 
bacteria~\cite{KeRiFeBo02,KiRi04,HiFuPaPe10,Trosvik10}, coral reef environments~\cite{BuJa79}, 
competing grasses~\cite{Watt47,Thorhallsdottir90,SiLiDa94}, etc. 
The simplest and most studied case corresponds to the Rock-Scissors-Paper (RSP) 
game, with ${\cal S}=3$, in which each strategy dominates the next one, in a
cyclic way~\cite{Gilpin75,Tainaka88}. These interactions, or food chain, are thus given by 
a three vertices, single looped oriented graph. Since there is no perfect ranking of 
the species, the system is fully intransitive.
A direct generalization~\cite{FrKrBe96,FrKr98,SaYoKo02,CaDuPlZi10,DuCaPlZi11} 
is to consider ${\cal S}>3$ competitors whose interactions also follow
an oriented ring, $0\to 1\to \ldots \to {\cal S}-1\to 0$.
% 
%On a lattice... for which general results on the fixation probability of a given species 
%exist~\cite{FrKr98,SaYoKo02} and the behavior may differ for even or odd $S$~\cite{SaYoKo02,DuCaPlZi11}. 
For the specific case of ${\cal S}=4$~\cite{SaYoKo02,SzSz04b,Szabo05,SzSz08,CaDuPlZi10,DoFr12,DuCaPlZi11,RoKoPl12}, 
the minimum value for which neutral pairs may exist, those non interacting  
alliances help prevent invasions. Such defensive alliances may also appear between
non mutually neutral species (cyclic alliances) when the interaction graph has more than a single 
loop~\cite{SiHoJoDa92,DuLe98,SzCz01a,SzCz01b,SzSz04b,Szabo05,PeSzSz07,SzSzSz07,SzSz08,SzSzBo08,LaSc08,LaSc09,HaPaKi09,LiDoYa12,AvBaLoMe12,AvBaLoMeOl12,RoKoPl12}.
Random and non regular food webs have also been considered~\cite{AbZa98,MaMiSnTr11,PaZlScCa11}.

Such models, with simplified competing interactions and food webs, do not
claim quantitative predictions, but attempt instead to unveil the
universal behavior that results from the direct competition between
species. The interactions are coarse grained in the sense that the
ultimate mechanism (dispute for space, resources, mating partners, etc) and its
non all-nothing nature (e.g., dependence on size, age, distance and other
contingent factors)
are averaged out and replaced by a simple, probabilistic interaction.
Such interactions may depend on space, time, be a characteristic of the
two species involved, etc., what introduces heterogeneities in the 
system~\cite{DuLe98,FrAb01,SaYoKo02,ClTr08,Masuda08,HeMoTa10,VePl10,CaDuPlZi10,DuCaPlZi11,JiZhPeWa11}. In turn, this gives rise to
hierarchical alliances and diverse levels of intransitivity.
Anomalous, negative responses may occur in this case, an example being the ``survival of the weakest''
principle, observed for ${\cal S}=3$~\cite{Tainaka93,FrAb01} 
(and its generalization for ${\cal S}>3$~\cite{CaDuPlZi10,DuCaPlZi11}), 
in which a species density may increase after its
invasion capability has been decreased. Real systems, with their more complex trophic networks, 
may even have more complex responses to variations in the invasion rates and, consequently, 
predicting their behavior in such a situation will be far more difficult.
%``the prey of the prey of the weakest is most likely to go extinct first''~\cite{CaDuPlZi10,DuCaPlZi11}.

Spatial correlations may exist when the range of interaction is limited but
play no role when the interactions are spatially unconstrained (fully mixed case), 
and simple mean field approximations are expected to produce reasonable results for
sufficiently large systems in such a case. Nonetheless, stochastic fluctuations are expected 
to become important for finite size populations and even drive the system
towards one of its absorbing states, in which one or more species become
extinct, decreasing the diversity. 

Intransitivity is considered a key mechanism for diversity sustaining in the
presence of competition. Thus, important questions arise on the effects of 
tuning the transitivity by changing the invasion probabilities. For example,  
does the diversity suddenly decrease once the system is no longer fully intransitive?  
%or a minimum amount of it is necessary before coexistence start decreasing?
Can diversity be predicted solely based on the structure of the interaction
graph? How does the system respond to changes on the interaction
parameters of a complex trophic network?

%\section{Model}

\begin{figure}[h] 
\includegraphics[width=8.5cm]{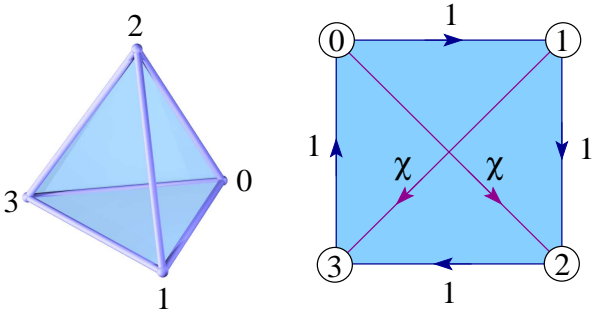}
\caption{Species interactions represented by an oriented graph (right), the arrows indicating the 
invasion direction and 1 and $0\leq \chi\leq 1$ are the corresponding rates. The possible
configurations, $(\rho_0,\rho_1,\rho_2,\rho_3)$, can be represented as points in a 3-simplex (left).
Upon extinction of a single species the 
orbits are constrained to one of the faces of the tetrahedron, a 2-simplex. After a second extinction, the
configurations are constrained to the line joining the two surviving species (1-simplex). If eventually a single species
remains, this state is represented by the corresponding vertex.}
\label{fig.4}
\end{figure}

To answer these questions, we start with a fully intransitive ring of four species 
(${\cal S}=4$) competing with the same unity invasion rate. All four species have similar roles,
with one prey and one predator each. 
This symmetry is broken when the 
interaction graph is turned into a fully connected graph, with
two diagonal interactions having a rate $\chi$ of invasion,
as  shown in Fig.~\ref{fig.4}. This introduces some hierarchy in
the system: the top species, 0 and 1, have two preys each (and
one predator), while the
bottom ones, 2 and 3, have two predators (and one prey). The arrows indicate
the direction in which the invasion occurs and the corresponding rate: around the original ring, invasions
occur with unitary rate while along the diagonals, this probability
is $\chi$. Species 2 and 3 have only one prey each,
but once they encounter their prey, they always subjugate them. On
the other hand, species 0 and 1 have two preys each, but with a
smaller than unity success rate and are the weakest species in this case
(see Section~\ref{section.conclusions} for the detailed discussion). 
When $\chi=1$ we recover the case considered in 
Ref.~\cite{LiDoYa12} (see also Ref.~\cite{Szabo05}).

The paper is organized as follows. Next section discusses the mean field approach for
the fully mixed version of the model, in particular, the stable fixed points both for
$\chi=0$ and $\chi\neq 0$. Then, we present the results for the spatially structured
system, with emphasis on the long term persistence of the coexistence state. Finally,
we discuss the similarities and discrepancies of both approaches and present our
conclusions.
 
\section{Analytical Results}

When spatial correlations are neglected and individuals have the
same probability to interact with all others, irrespective of
their distance, one may attempt a mean field description. 
Let $\rho_i$ be the density of species $i$
(obviously $\sum_i\rho_i=1$). Time variations in the densities
may only occur due to interactions between different species,
in which the stronger one will invade the weaker with rate
1 or $\chi$. The mean field equations depend only on the
frequency of such encounters and read:
\begin{equation}
\dot{\rho_i} = \sum_j I_{ij}\rho_i\rho_j ,
\label{eq.mf}
\end{equation}
where each element of the interaction matrix, $I_{ij}$, is the rate with
which species $i$ invades $j$. A negative $I_{ij}$ means that the invasion 
direction is reversed.  The matrix ${\bm I}$ is given by
\begin{equation}
{\bm I} = \left[\begin{array}{cccc}
0 & 1 & \chi & -1\\
-1 & 0 & 1 & \chi\\
-\chi & -1 & 0 & 1\\
1 & -\chi & -1 & 0
\end{array}\right].
\label{eq.G0}
\end{equation}
These equations present several equilibrium points such that $\dot{\rho}_i=0, \forall i$. The 
linear stability of these steady states is determined by the sign of the real part of
the eigenvalues of the Jacobian matrix. If at least one eigenvalue has a positive
real part, the corresponding fixed point is unstable, otherwise it is stable. 
Furthermore, a stable equilibrium point may be asymptotically attainable
when all real parts are strictly negative. When there are purely imaginary eigenvalues,
the stable equilibrium is neutral and never attainable dynamically. 

The fixed points for $\chi=0$ have been discussed by several 
authors~\cite{SaYoKo02,SzFa07,DoFr12,CaDuPlZi10,DuCaPlZi11}. First,
there are four absorbing states that are heteroclinic points (saddle points)~\cite{HoSi98}, 
at the vertices of the 3-simplex, 
in which only one species survives: $(1,0,0,0)$, $(0,1,0,0)$, $(0,0,1,0)$ 
or $(0,0,0,1)$. 
In addition to these, and because species 0 and 2 (or 1 and 3) are mutually neutral, 
any point on the line connecting each pair is also a fixed point, the initial proportion 
between them kept constant:
\begin{eqnarray}
&& \left(c_o,0,1-c_o,0\right) \label{eq.c0a} \\
&& \left(0,c_o,0,1-c_o\right) \label{eq.c0b},
\end{eqnarray}
with $0\leq c_o\leq 1$. 
Lastly, there is a coexistence fixed point in the interior of the 3-simplex, 
for which all densities are non zero, 
\begin{equation}
\left(c_o,\frac{1}{2}-c_o,c_o,\frac{1}{2}-c_o\right),
\label{eq.coex}
\end{equation}
with $0\leq c_o\leq 1/2$, a particular example being the symmetric state $(1/4,1/4,1/4,1/4)$.
This point is stable, but not asymptotically stable. In fact, there are two integrals of motion: $\rho_0 \rho_2$ 
and $\rho_1 \rho_3$. In contrast, for the ${\cal S}=3$ game there is only one invariant of motion: $\rho_0 \rho_1 \rho_2$.

\begin{figure}[hbt]
\begin{center}
\includegraphics[width=8.5cm]{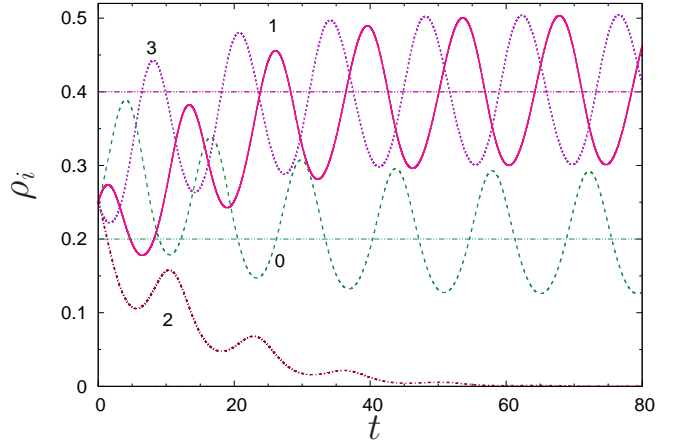}
\caption{Example of the time evolution for $\chi=0.5$ within mean field, starting with
$\rho_i(t=0)=1/4, \forall i$. Species 2 rapidly becomes extinct and the remaining three 
species oscillate,
out of phase, around the respective neutrally stable fixed point (dashed lines) given
by Eq.~(\ref{eq.eq1}), $(0.2,0.4,0,0.4)$. The fixed point also corresponds  to an
average of $\rho_i$ over a period, once the stationary state
is attained.}
\label{fig.dXt_top0}
\end{center}
\end{figure}

With $\chi\neq 0$, the coexistence state Eq.~(\ref{eq.coex}) is no longer a solution
of Eq.~(\ref{eq.mf}).
Nonetheless,  
there are two further fixed points in which one species (1 or 2) 
dies out and the remaining three species form a non homogeneous RSP game~\cite{LiDoYa12}:
\begin{eqnarray}
&& \left(\frac{1}{2+\chi},0,\frac{1}{2+\chi},\frac{\chi}{2+\chi}\right) \\
&& \left(\frac{\chi}{2+\chi},\frac{1}{2+\chi},0,\frac{1}{2+\chi}\right). \label{eq.eq1}
\end{eqnarray}
Notice that for $\chi=0$ the above fixed points are particular cases of Eqs.~(\ref{eq.c0a}) and (\ref{eq.c0b}),
respectively.
The first solution, in which the species 1 becomes
extinct, is an unstable fixed point, while the second one, in which  
species 2 goes extinct and the remaining three compose a heterogeneous RSP game, is 
(neutrally) stable. In the limit $\chi\to\infty$, the stable solution  Eq.~(\ref{eq.eq1})
becomes $(1,0,0,0)$ and species 0 dominates.
An example, Fig.~\ref{fig.dXt_top0}, shows the evolution from
the symmetrical initial state with $\rho_i=1/4, \forall i$ and $\chi=0.5$.
The system
approaches a closed orbit that oscillates around $(0.2,0.4,0,0.4)$, Eq.~(\ref{eq.eq1}), 
after the exponentially fast extinction of species 2. When $\rho_2=0$, the quantity 
$\rho_0^{\chi}\rho_1\rho_3$ is an integral of motion~\cite{HoSi98,IfBe03,ReMoFr06}. Interestingly,
besides the trivial normalization condition, no invariant involving all four densities exists for $\chi\neq 0$.  The
fixed points are equivalent to the time average of the oscillating densities. 
Both the period of the oscillations and the time that species
2 takes to become extinct diverge when $\chi\to 0$ since 
in this limit the existence of the invariants of motion mentioned above precludes the possibility of an extinction.
Indeed, when $\chi=0$, the homogeneous initial condition considered here becomes a 
fixed point with four coexisting species.
For the homogeneous case, $\chi=1$, one recovers the $\rho_i=1/3, \forall i$,
solution.
Notice that each species density depends on its prey's invasion rate and when we decrease
$\chi$ (species 1 invasion rate over 3), although one would expect a decrease in its density, 
it is the density of its predator, species 0, instead, that decreases. This is known
as the ``survival of the weakest'' principle~\cite{Tainaka93,FrAb01}.

When spatial correlations are important, as when agents are placed on a lattice
(see Section~\ref{section.simulation}), the mean field approach usually breaks down.
One of the simplest ways to go beyond the mean field predictions is to use the pair 
approximation (PA)~\cite{MaDi99}. Within this approach one considers the dynamics of 
pairs of connected sites (instead of only one-site quantities as in MF). As the 
corresponding equations depend on triplets of connected sites, the system is closed 
by choosing an ansatz relating three- and two-site quantities. For the PA the ansatz 
chosen is of the form $P(123)=P(12) P(23)/P(2)$, where $P(123)$ is the probability of 
having species 1, 2, and 3 occupying three connected sites (2 occupies the central site). 
$P(12)$ and $P(23)$ are similarly defined. 

\begin{figure}[htb]
\begin{center}
\includegraphics[width=8.5cm]{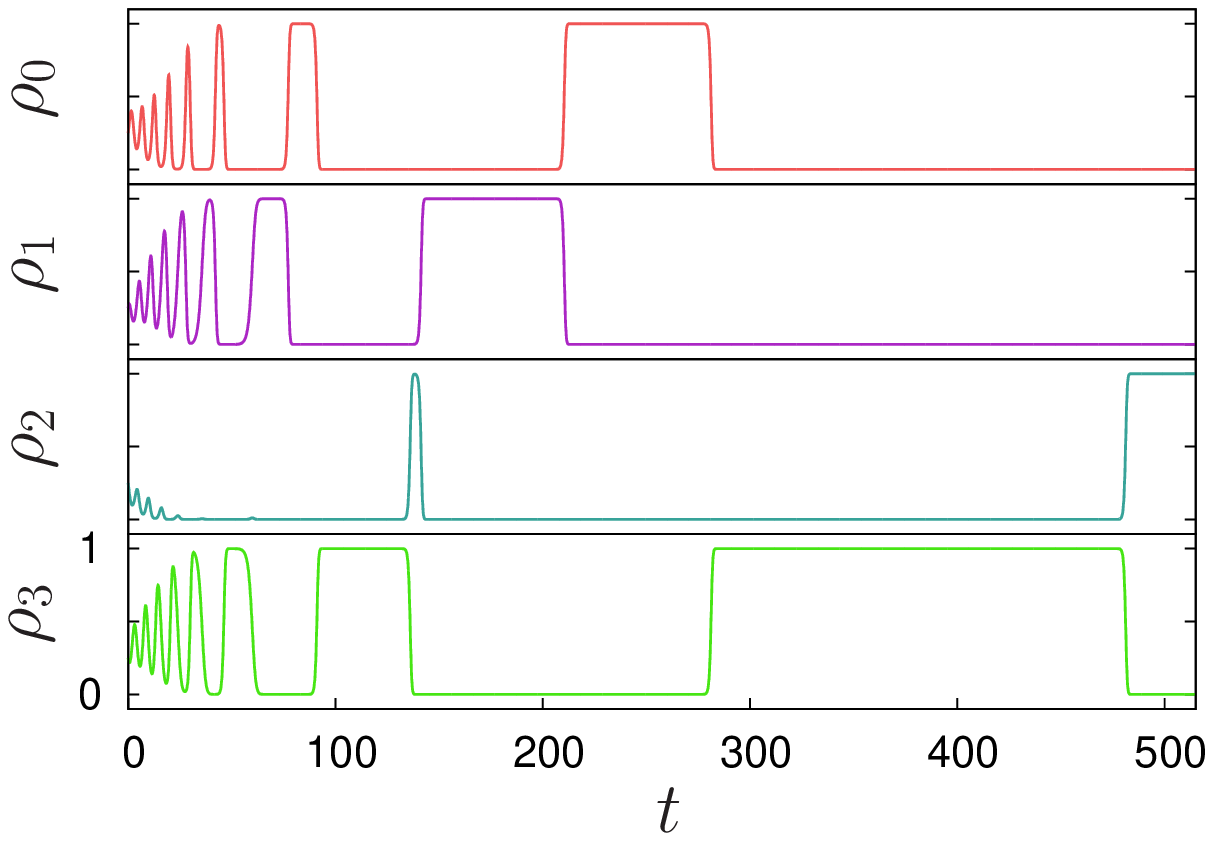}

\includegraphics[width=6cm]{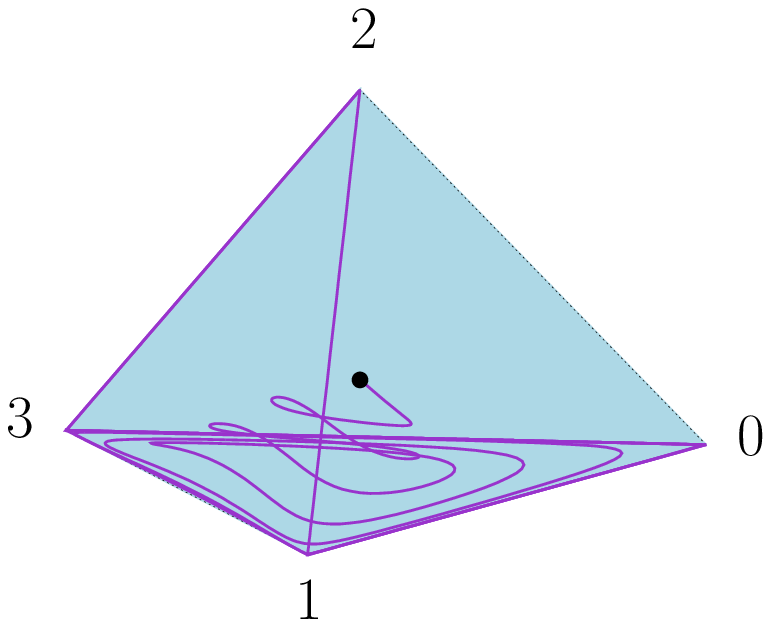}
\caption{(Top panel) Numerical resolution of the pair approximation equations for
$\chi=0.45$, starting with
$\rho_i(t=0)=1/4, \forall i$. Notice that $\rho_2$ decreases to very small values during
the transient, and the system approaches the 310 cycle. Later, however, $\rho_2$ grows once 
again and the system switches to the 3210 cycle. (Bottom panel) The same orbit plotted
in the simplex. The black dot signals the starting point. Although initially the orbit 
approaches the 310 face and the species 2 seems to go extinct, it eventually resumes and
the full 3210 cycle is populated (notice that the orbit goes very close to the
edges of the simplex at later times).}
\label{fig.pairapp}
\end{center}
\end{figure}

For the system considered here the PA does not have any fixed points with coexistence 
of the four species. In fact, when expressed in term of species densities, the fixed points 
of the PA coincide with those found using MF. One important difference is that the fixed point 
for which there is extinction of species 2 is asymptotically unstable. As happens in the case 
of the RSP game~\cite{MaLe75,SzSzIz04}, the model has a heteroclinic cycle involving 
the four species. In addition, the diagonal interactions give rise to two new heteroclinic 
cycles involving species 0, 1 and 3, and 0, 2 and 3. As these cycles share some nodes, none of 
them can be asymptotically stable. This, however, does not mean that they cannot dominate the 
dynamics. Solving the PA equations of motion for several different initial conditions shows that 
for long times the system asymptotically approaches the 3210 cycle. However, for shorter times the 
system moves towards the vicinity of the 310 cycle, and can stay there for a long time until it 
``jumps'' to the vicinity of the 3210 cycle. This can be thought of as a `competition' between the 
cycles~\cite{KiSi94} that is eventually won by the cycle 3210 (the cycle 023 does not seem to play 
any role in the dynamics). The fact that the density of species 2 falls to extremely low levels during 
the transient implies that in a stochastic version of this dynamics the extinction of species 2 would 
happens after a rather short time. The duration of this transient is an increasing function of $\chi$. 
Unfortunately, it is not possible to obtain the time average of the densities of any of the species 
because these quantities do not converge~\cite{Ga92}. The above behavior is illustrated in
Fig.~\ref{fig.pairapp}. In particular, notice that species 2 resumes to a noticeable density in the
top panel of Fig.~\ref{fig.pairapp}. All four densities appear on the heteroclinic orbit, one
at a time (with the others being extremely small) with increasing periods of stasis on each
(unstable) monoculture~\cite{MaLe75,HuWe01}. Although all four species have non zero 
densities, the heteroclinic cycle is termed ``impermanent coexistence''~\cite{HuLa85} since
they do not coexist with finite densities.
In the bottom panel, on the other hand, the orbit is
depicted in the simplex of Fig.~\ref{fig.4}. Initially (the starting point is the black dot)
the orbit approaches the 310 face but eventually species 2, that was only apparently extinct,
increases its density once again and the system stabilizes on the full 3210 heteroclinic cycle.

To summarize,  the system tends to decrease the amount of hierarchy (because of the
$\chi$ weighted connections) by the exponentially fast extinction of one species, converging
to a fully intransitive, non hierarchical, three species system~\cite{LiDoYa12}. 
Indeed, in mean field, any amount of transitivity (measured by $\chi$) destroys the
possible coexistence state that exists when $\chi=0$.

\section{Simulations}  
\label{section.simulation}
  
The dynamics on a lattice may be very different from the evolution
predicted by the mean field equations, mainly because the range of 
interaction being much smaller than the system size, local 
correlations play an important role. Moreover, unless the system
is very large, finite size effects exist and introduce stochastic
effects. As an example, the invariants
discussed in the previous section, quantities that are kept constant
during the motion along closed orbits, no longer persist for finite
systems, and density fluctuations eventually drive the system, through
extinctions, into an absorbing state. These finite size effects become less
important for large systems and disappear for $L\to\infty$,
where $L$ is the system linear size. In order to study the system on
the lattice, we consider a square grid with $N=L^2$ sites with periodic
boundary conditions and, with the same probability, one of the four species is randomly assigned to
each of those sites at $t=0$. One site and one of its neighbors is
chosen at random and the stronger site invades the other, depending on the species, 
with probability either unity or $\chi$. This step is
repeated $N$ times, what defines the time unit. Analogous to the
mean field approach, the densities oscillate in time, however, the
amplitude of these oscillations seems to decrease with the size of
the system and tend to disappear for very large systems.

\begin{figure}[htb]
\begin{center}
\includegraphics[width=8.5cm]{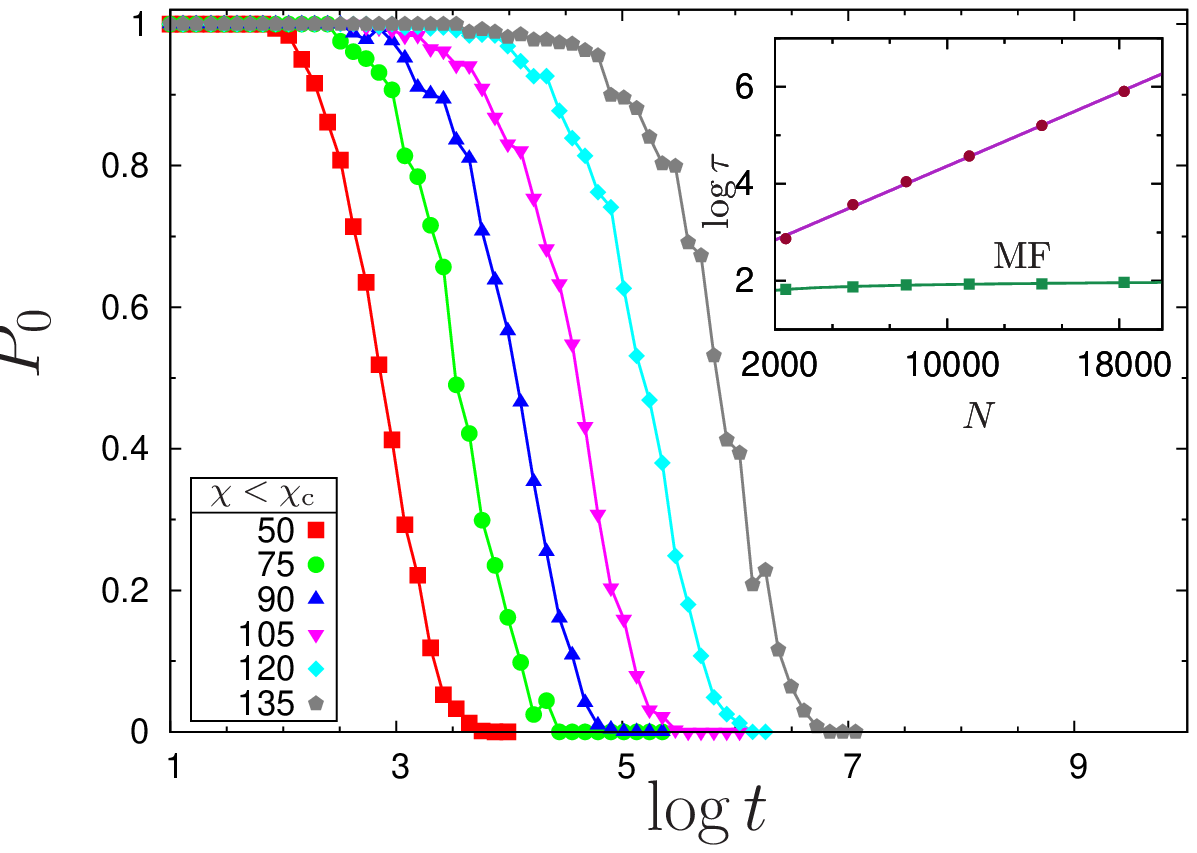}

\includegraphics[width=8.5cm]{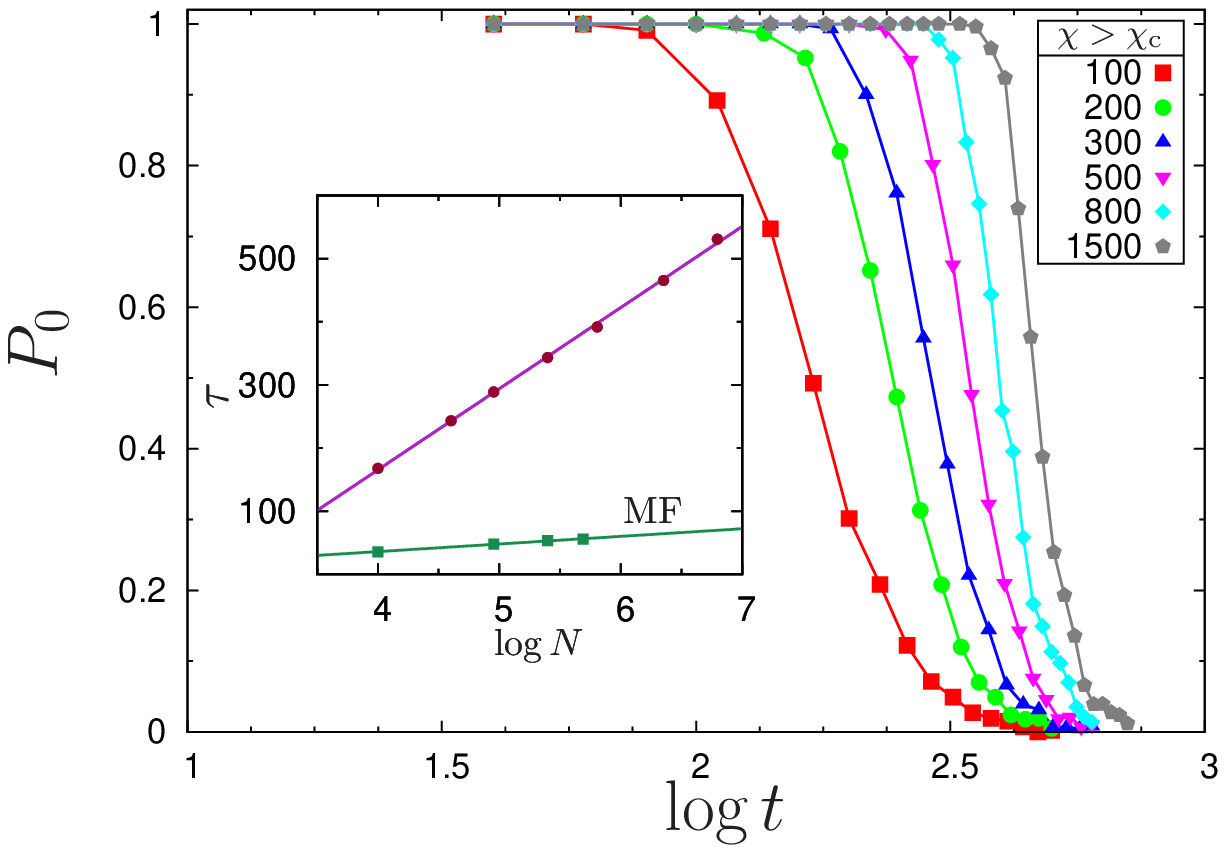}
\caption{(Top panel) Probability of all species surviving up to time $t$, for 
$\chi=0.31<\chic$ and 
several linear sizes $L$. Data are averages over at least 200 samples. The characteristic
extinction time $\tau$, defined as $P_0(\tau)\equiv 1/2$, is shown in the inset and grows
exponentially with the system size. For comparison, we also
show in the inset the characteristic extinction time for a finite fully mixed system (see
text). (Bottom panel) The same, but for 
$\chi=0.5>\chic$. The characteristic extinction time shown in the inset has a logarithmic 
dependence on $N$. Notice the very different horizontal scales in the two panels.}
\label{fig.P0}
\end{center}
\end{figure}

\begin{figure}[htb]
\begin{center}
\includegraphics[width=8.5cm]{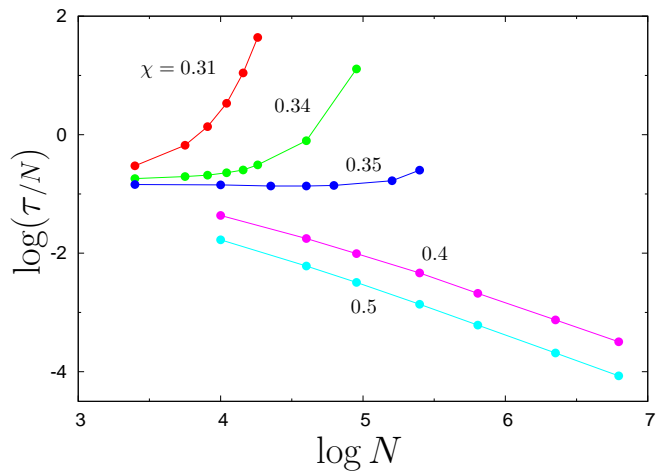}
\caption{Average extinction time $\tau$ versus $N$ for several values of $\chi$. 
The critical value of $\chi$, $\chic$, is slightly above 
$\chi=0.35$. Notice the two different asymptotic behaviors, $\tau(\chi<\chic)/N\to\infty$ and
$\tau(\chi>\chic)/N\to 0$. For $\chi\simeq\chic$, $\tau$ is linear in $N$.  
The lines are only guide to the eyes. }
\label{fig.tau}
\end{center}
\end{figure}

Even though a deterministic system may have stable coexistence states, in its stochastic 
counterpart the finite number of interacting agents induce fluctuations that, given enough 
time, eventually lead to the extinction of one or more species.
%Whatever the value of $\chi$, a finite system is always driven away from the coexistence
%phase by a density fluctuation. 
However, as the system size increases, distinct dynamical 
behaviors may be observed depending on the value of $\chi$. The dependence of the average 
characteristic time for an extinction to occur on the system size $N$ allows for a
classification of the possible occurrying scenarios~\cite{AnSc06,ReMoFr07a,CrReFr09b,Frey10}.
The coexistence is said to be stable when the related deterministic dynamics presents
a stable atractor in the coexistence phase, and this is associated with an exponentially
increasing time for the first extinction to occur as $N$ increases.  Analogously, the
unstable state presents a logarithmic increase of the extinction time and the deterministic
system approaches an absorbing state. In between, a power law dependence of the extinction
time on the system size is related with the presence of closed, neutrally stable orbits
in the deterministic case. 
The top panel of Fig.~\ref{fig.P0} shows, for small values of $\chi$ and
several linear sizes $L$, the probability that the system does 
not suffer any extinction up to the time $t$, $P_0(t)$, that is, the probability 
of a persistent coexisting state. The larger the system is, the longer
it takes for $P_0$ to start dropping. 
We may define a characteristic time for the first 
extinction, $\tau(N)$, as the time when $P_0$ drops to half its
initial value, that is, $P_0(\tau)\equiv 1/2$. In the inset of Fig.~\ref{fig.P0}, top
panel, one can observe, for the range of sizes considered here, 
that $\tau(N)$ has an exponential growth 
and even for modest sizes, the time of the first extinction is very
large. Extinction~\cite{OvMe10} in this case is driven by very rare fluctuations
and the coexistence is said to be stable~\cite{AnSc06,ReMoFr07a,CrReFr09b,Frey10}.
On the other hand, for large $\chi$, inset of Fig.~\ref{fig.P0},
bottom panel,
the extinction time growth is logarithmic in $N$ and even for very large
systems (one order of magnitude larger than in the previous case), $\tau$
is rather small. Coexistence in this case is unstable and even small
fluctuations are able to drive some species to extinction~\cite{ReMoFr07a,CrReFr09b}.
Thus, comparing these two cases, there must be a dynamical 
critical value of $\chi$, $\chic$, separating those two quite distinct dynamical 
behaviors of $\tau(N)$, and a rough estimate places this critical value at
$\chic\simeq 0.35$. Indeed, as shown in Figs.~\ref{fig.P0} and \ref{fig.tau}, 
$\tau(\chi<\chic)$ and $\tau(\chi>\chic)$ have very distinct asymptotic behavior~\cite{ScCl10}.
While for $\chi<\chic$ the mean extinction time $\tau$ grows exponentially,
above $\chic$ this growth is logarithmic in $N$. The intermediate region,
for $\chi\simeq\chic$, the scaling of $\tau$ with the system size is
polynomial. %, $\tau\sim N^2$. 
 For $\chi>\chic$ species 2 goes extinct and the
three remaining ones converge to densities close to the fixed
point Eq.~(\ref{eq.eq1}). It is also important to stress that a second extinction,
when it occurs, takes a much longer timescale. Both insets of Fig.~\ref{fig.P0}
also show the comparison with the correspondent $\tau$ for a finite fully
mixed system. To simulate such a system, in each MC step, new neighbors are randomly 
assigned to each site, without any distance constraint. For both $\chi>\chic$ and
$\chi<\chic$,  $\tau$ grows logarithmicly with $L$, but with a 
small declivity, and no distinction exists between the two regions.

\begin{figure}[htb]
\begin{center}
\includegraphics[width=8.5cm]{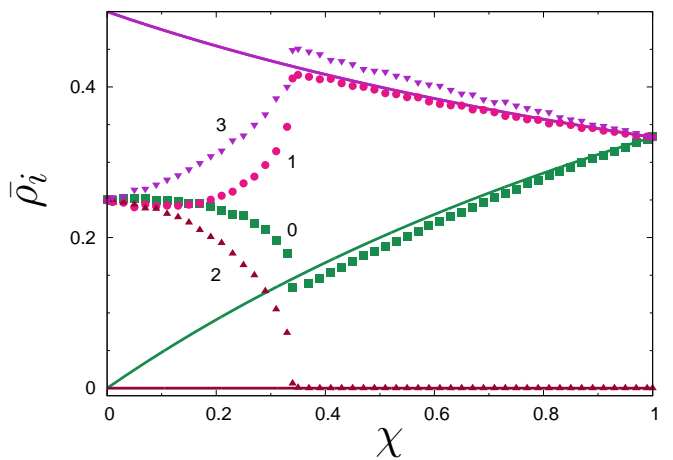}
\caption{Stationary densities as a function of $\chi$. Although on a lattice we do not observe
an oscillating behavior, there are sample to sample fluctuations and we  average over
at least 1000 samples (symbols). The lines are the fixed point, Eq.~(\ref{eq.eq1}), of the mean field equations.
Notice that albeit the reasonable agreement for $\chi>\chic$, below this value the system
is in a state, not captured by the mean field approach, in which all four species
coexist.}
\label{fig.densities}
\end{center}
\end{figure}

For $\chi<\chic$, due to the exponential growth of $\tau$, the coexistence state is
said to be stable and large systems stay in a state in which all four species attain a non zero
fixed point. The average asymptotic density $\overline{\rho}_i$ for
large systems can be obtained by extrapolating the above behavior, that is,
$\overline{\rho}_i = \lim_{t\to\infty}\lim_{N\to\infty} \rho_i(t)$ (notice
that the limits are not interchangeable). The results are shown in
Fig.~\ref{fig.densities} as a function of $\chi$. The dynamical transition
is clearly seen as the point at which species 2 goes extinct. Notice that all four
densities are different, both above and below $\chic$.

Comparing with the mean field predictions, we notice several
fundamental differences. First of all, the coexistence state
only appears on the regular lattice since for $\chi\neq 0$,
in mean field, species 2 always gets extinct. Thus, the
dynamical transition that we observe does not exist for
a fully mixed system. Indeed, we can
simulate the system with annealed neighbors, and the characteristic
time of the first extinction is small, presenting a logarithmic
growth with $N$ for all $\chi$,
as can be seen in both insets of Fig.~\ref{fig.P0}. 
The second difference is that although oscillations are observed
for finite systems, they tend to disappear for very large sizes~\cite{LaSc09}. 
A possible explanation is that for 
a sufficiently large system, several different regions will
evolve almost independently with uncorrelated phases, such that
the overall system no longer presents oscillations.
The third difference is that while the mean field behavior
is monotonous on $\chi$ ($\overline{\rho}_2$ is always zero, $\overline{\rho}_0$ is
always increasing and $\overline{\rho}_1=\overline{\rho}_3$ is always decreasing), 
on the regular lattice the behavior is non monotonous, 
Fig.~\ref{fig.densities}. Species 1 and 3 densities,
now resolved, increase for $\chi$ up to $\chic$ and
decrease afterwards (species 0 has the opposite behavior).
Notice, however, that although above $\chic$ the agreement with 
mean field  is reasonable, below $\chic$ there is both
qualitative and quantitative disagreement. In particular, all
four densities are different and no fixed point predicted by
the mean field approach has such a property for $\chi\neq 0$.

%A more sophisticated
%analysis must be able to predict the observed dependence on $\chi$
%as well.

\section{Conclusions}
\label{section.conclusions}

We studied a minimal model for a trophic network presenting multiple
loops of interacting species, focusing on the effects of a tunable
transitivity on the persistence of the coexistence state. As the invasion
rate $\chi$ is changed, we observed two distinct dynamical phases separated by
a transition at $\chic\simeq 0.35$, one for $\chi<\chic$ in which the
coexistence state is stable (the mean extinction time exponentially grows
with the size of the system) and the other for $\chi>\chic$ in which one
species goes extinct on logarithmic timescales and the system ends up
performing a heterogeneous RSP game. At the transition region between
those two regimes, $\chi\simeq\chic$, $\tau$ presents a polynomial scaling
with the size of the system. This transition, and the coexistence state observed 
in the simulations are not captured by the mean field approach.

For $\chi=0$, each species around the external four species loop
has one prey and one predator. In addition, for $\chi\neq 0$ there are 
four internal loops with three species, two intransitive (013 and 023) 
and two transitive (012 and 123). In this case, because of the even 
number of species, the number of predators and preys of 
each species may differ. Thus, depending on the arrows
orientation, there are three possible choices for the number of preys
(or, equivalently, predators): $(3,2,1,0)$, $(3,1,1,1)$ and $(2,2,1,1)$.
We only considered the last structure, Fig.~\ref{fig.4}, that is somewhat intermediate between 
an intransitive and a hierarchical system.  
The larger is $\chi$, the less intransitive the system is and
one can expect that the amount of coexistence will decrease.
However, we have shown that under the presence of spatial correlations and
a not too large transitivity, this system may persist in a state
of full diversity over exponentially large timescales. For larger levels
of transitivity, on the other hand, the system eventually evolves into a three
species hierarchical system, irrespective of the spatial structure~\cite{LiDoYa12}. 
We thus observe a dynamical transition between these two regimes on the spatially 
structured system, at $\chic$, not captured by a mean field analysis. Notice that any
extinction drives the system into an absorbing state and diversity, due to the
absence of mutations, is an always
decreasing quantity for this class of model.
Below $\chic$, mean field is not a good approximation for the lattice dynamics 
of our system either quantitatively or qualitatively. 
The threshold value of $\chi$ also indicates that above a
certain level of intransitivity ($\chi>\chic$), spatial correlations are no longer important
and the system on a lattice is attracted to the mean field fixed points 
(the densities are non oscillating). Even though the pair approximation is assumed to be a better approximation than mean field, in our case it does not provide a better description of the dynamics, in terms of fixed points. For short times the dynamics of the PA does look similar to the mean field dynamics, since for large values of $\chi$ the density of species 2 drops to extremely low values, but for longer times the dynamics is dominated by a heteroclinic cycle involving all four species.
%Nonetheless,
%even this observed agreement between the simulation and the mean field calculation 
%for $\chi>\chic$ is surprising, since the specific mechanism by which higher
%levels of transitivity decrease the effect of the correlations responsible for the coexistence
%state is not clear. 

Non monotonous responses driven by the spatial correlations are observed, while
the mean field approach predicts a monotonous behavior as $\chi$ changes.  
For $\chi<\chic$, species 2 and 3 (and, analogously, 0 and 1) respond 
in opposite ways: while $\overline{\rho}_3$ increases with $\chi$, $\overline{\rho}_2$ decreases. 
The opposite behavior was predicted in the mean field approach. On the
other hand, above $\chi_c$ the trends agree with the mean field prediction. 
Interestingly, in spite of presenting opposite behavior when $\chi$ increases, 
species 0 and 1 both become more aggressive. Since both predate on 2, this species 
has the smallest density (and becomes extinct in mean field).
For $\chi>\chic$, $\overline{\rho}_2=0$ and the remaining three species form a heterogeneous
RSP game that obeys, both on the lattice and in MF, the usual 
``survival of the weakest'' principle: as the invasion rate $\chi$ of the
weakest species (1) increases, its density $\overline{\rho}_1$ decreases,
while the density of ``the prey of the prey of the weakest''~\cite{DuCaPlZi11}, in
this case species 0, increases. For $\chi<\chic$, since all four
species survive, the  ``the prey of the prey of the weakest'' principle~\cite{DuCaPlZi11} 
must be modified
because some species have multiple preys. Although species 0 and 1
have a wider range of possible targets, they are less efficient since their 
overall success rate is less than 1 ($1+\chi\leq 2$), and may be considered the
weakest species. Species 2 and 3, on the
other hand, fully overtake their preys. Nonetheless,  the prey of the
two weakest (species 2), itself stronger than them, goes extinct. 
Thus, although there is no obvious generalization of the above principle, one notice that
the ambiguity in defining strong and weak in this case may be raised by allowing all six 
parameters to be different, what may, in turn, allow for such an statement.   
%since some
%of the species have multiple preys, it can be stated as {\it the weakest
%preys (0 and 2) of the prey of the weakest are the least like to survive}, whatever
%is the value of $\chi$. 
It is also clear that statements like this will
become more intricate as the number of species increases.

Further questions arise for such systems. For example, in order to better 
understand the effects of different levels of transitivity, in particular to probe
anomalous responses as the ``survival of the weakest'', the study of
other trophic structures with four species, and larger values of ${\cal S}$ as well, is important. 
In addition,
finite populations may have a different behavior. Indeed, the community size, besides 
setting the scale for the average extinction time, may also influence which is the surviving 
species~\cite{MuGa10}. 
% se d=1 for trivial, comentar no primeiro paragrafo das simulacoes
One may also probe the robustness of the results presented here, for
example, by studying different lattices (random graph, small world, etc), dimensions
and initial conditions. On a regular lattice, geometric and dynamical properties of
the evolving groups are also of interest~\cite{AvBaLoMe12,AvBaLoMeOl12}.
How to properly quantify the transitivity of a trophic network and correlate
it with the coexistence present in a population is still an open problem
(e.g., \cite{Petraitis79,LaSc06,LaSc08,LaSc09,RoAl11,ShMc12} and references therein).
We have shown that the structure alone is not enough to predict whether there will
be coexistence or not. Considering the trophic relations as a weighted
network may lead to an index allowing different levels of coexistence based
on the same structure. Finally, allowing general weights on the trophic
network~\cite{DuCaPlZi11}
shall present an even richer behavior in the presence of crossed interactions.

%\begin{acknowledgments}
\chapter*{Acknowledgements}
The authors acknowledge partial support from the Brazilian agency CNPq,
in particular through the PROSUL 490440/2007 project.  JJA is
partially supported by Brazilian agencies Fapergs and CAPES (BEX 0698/11 and
CAPES/COFECUB 667/10),   and
is a member of the INCT Sistemas Complexos.
%\end{acknowledgments}

\bibliographystyle{elsarticle-num}
%\bibliography{rsp4}

\begin{thebibliography}{10}
\expandafter\ifx\csname url\endcsname\relax
  \def\url#1{\texttt{#1}}\fi
\expandafter\ifx\csname urlprefix\endcsname\relax\def\urlprefix{URL }\fi
\expandafter\ifx\csname href\endcsname\relax
  \def\href#1#2{#2} \def\path#1{#1}\fi

\bibitem{HoSi98}
J.~Hofbauer, K.~Sigmund, Evolutionary Games and Population Dynamics, Cambridge
  University Press, Cambridge, 1998.

\bibitem{SzFa07}
G.~Szabó, G.~Fath, Evolutionary games on graphs, Phys. Rep. 446 (2007)
  97--216.

\bibitem{Frey10}
E.~Frey, Evolutionary game theory: Theoretical concepts and applications to
  microbial communities, Physica A 389 (2010) 4265--4298.

\bibitem{SiLi96}
B.~Sinervo, C.~Lively, The rock-paper-scissors game and the evolution of
  alternative male strategies, Nature 380 (1996) 240--243.

\bibitem{KeRiFeBo02}
B.~Kerr, M.~A. Riley, M.~W. Feldman, B.~J.~M. Bohannan, Local dispersal
  promotes biodiversity in a real-life game of rock–paper–scissors, Nature
  418 (2002) 171--174.

\bibitem{KiRi04}
B.~C. Kirkup, M.~A. Riley, Antibiotic-mediated antagonism leads to a bacterial
  game of rock-paper-scissors {\it in vivo}, Nature 428 (2004) 412--414.

\bibitem{HiFuPaPe10}
M.~E. Hibbing, C.~Fuqua, M.~R. Parsek, S.~B. Peterson, Bacterial competition:
  surviving and thriving in the microbial jungle, Nature Reviews: Microbiology
  8 (2010) 15--25.

\bibitem{Trosvik10}
P.~Trosvik, K.~Rudi, K.~O. Strætkvern, K.~S. Jakobsen, T.~Næs, N.~C.
  Stenseth, Web of ecological interactions in an experimental gut microbiota,
  Environ. Microb. 12 (2010) 2677--2687.

\bibitem{BuJa79}
L.~W. Buss, J.~B.~C. Jackson, Competitive networks: Nontransitive competitive
  relationships in cryptic coral reef environments, Am. Nat. 113 (1979)
  223--234.

\bibitem{Watt47}
A.~S. Watt, Pattern and process in the plant community, J. Ecol. 35 (1947)
  1--22.

\bibitem{Thorhallsdottir90}
T.~E. Th\'orhallsd\'ottir, The dynamics of five grasses and white clover in a
  simulated mosaic sward, J. Ecol. 78 (1990) 909--923.

\bibitem{SiLiDa94}
J.~Silvertown, C.~E.~M. Lines, M.~P. Dale, Spatial competition between grasses
  -- rates of mutual invasion between four species and the interaction with
  grazing, J. Ecol. 82 (1994) 31--38.

\bibitem{Gilpin75}
M.~E. Gilpin, Limit cycles in competition communities, Am. Nat. 109 (1975)
  51--60.

\bibitem{Tainaka88}
K.-I. Tainaka, Lattice model for the {L}otka-{V}olterra system, J. Phys. Soc.
  Japan 57 (1988) 2588.

\bibitem{FrKrBe96}
L.~Frachebourg, P.~L. Krapivsky, E.~Ben-Naim, Spatial organization in cyclic
  lotka-volterra systems, Phys. Rev. E 54 (1996) 6186--6200.

\bibitem{FrKr98}
L.~Frachebourg, P.~L. Krapivsky, Fixation in a cyclic lotka - volterra model,
  J. Phys. A: Math. Gen. 31~(15) (1998) L287--L293.

\bibitem{SaYoKo02}
K.~Sato, N.~Yoshida, N.~Konno, Parity law for population dynamics of n-species
  with cyclic advantage competitions, Appl. Math. Comput. 126 (2002) 255--270.

\bibitem{CaDuPlZi10}
S.~O. Case, C.~H. Durney, M.~Pleimling, R.K.P.Zia, Cyclic competition of four
  species: Mean-field theory and stochastic evolution, EPL 92 (2010) 58003.

\bibitem{DuCaPlZi11}
C.~H. Durney, S.~O. Case, M.~Pleimling, R.~K. P.Zia, Saddles, arrows, and
  spirals: Deterministic trajectories in cyclic competition of four species,
  Phys. Rev. E 83 (2011) 051108.

\bibitem{SzSz04b}
G.~Szabó, G.~A. Sznaider, Phase transition and selection in a four-species
  cyclic predator-prey model, Phys. Rev. E 69 (2004) 031911.

\bibitem{Szabo05}
G.~Szabó, Competing associations in six-species predator-prey models, J. Phys.
  A: Math. Gen. 38 (2005) 6689--6702.

\bibitem{SzSz08}
G.~Szab\'o, A.~Szolnoki, Phase transitions induced by variations of invasion
  rates in spatial cyclic predator-prey models with four or six species, Phys.
  Rev. E 77 (2008) 011906.

\bibitem{DoFr12}
A.~Dobrinevski, E.~Frey, Extinction in neutrally stable stochastic
  {L}otka-{V}olterra models, Phys. Rev. E 85 (2012) 051903.

\bibitem{RoKoPl12}
A.~Roman, D.~Konrad, M.~Pleimling, Cyclic competition of four species: domains
  and interfaces, arXiv:1205.4914 (2012).

\bibitem{SiHoJoDa92}
J.~Silvertown, S.~Holtier, J.~Johnson, M.~P. Dale, Cellular automaton models of
  interspecific competition for space -- the effect of pattern on process, J.
  Ecol. 80 (1992) 527--534.

\bibitem{DuLe98}
R.~Durrett, S.~Levin, Spatial aspects of interspecific competition, Theor. Pop.
  Biol. 53 (1998) 30--43.

\bibitem{SzCz01a}
G.~Szabó, T.~Cz\'ar\'an, Defensive alliances in spatial models of cyclical
  population interactions, Phys. Rev. E 64 (2001) 042902.

\bibitem{SzCz01b}
G.~Szabó, T.~Cz\'ar\'an, Phase transition in a spatial lotka-volterra model,
  Phys. Rev. E 63 (2001) 061904.

\bibitem{PeSzSz07}
M.~Perc, A.~Szolnoki, G.~Szab\'{o}, Cyclical interactions with
  alliance-specific heterogeneous invasion rates, Physical Review E 75~(5)
  (2007) 052102.

\bibitem{SzSzSz07}
G.~Szabó, A.~Szolnoki, G.~A. Sznaider, Segregation process and phase
  transition in cyclic predator-prey models with an even number of species,
  Phys. Rev. E 76 (2007) 051921.

\bibitem{SzSzBo08}
G.~Szabó, A.~Szolnoki, I.~Borsos, Self-organizing patterns maintained by
  competing associations in a six-species predator-prey model, Phys. Rev. E 77
  (2008) 041919.

\bibitem{LaSc08}
R.~A. Laird, B.~S. Schamp, Does local competition increase the coexistence of
  species in intransitive networks?, Ecology 89 (2008) 237--247.

\bibitem{LaSc09}
R.~A. Laird, B.~S. Schamp, Species coexistence, intransitivitym, and
  topological variation in competitive tournaments, J. Theor. Biol. 256 (2009)
  90--95.

\bibitem{HaPaKi09}
S.-G. Han, S.-C. Park, B.~J. Kim, Reentrant phase transition in a predator-prey
  model, Phys. Rev. E 79 (2009) 066114.

\bibitem{LiDoYa12}
Y.~Li, L.~Dong, G.~Yang, The elimination of hierarchy in a completely cyclic
  competition system, Physica A 391 (2012) 125--131.

\bibitem{AvBaLoMe12}
P.~P. Avelino, D.~Bazeia, L.~Losano, J.~Menezes, Von-neumann's and related
  scaling laws in rock-paper-scissors type games, arXiv:1203.6671 (2012).

\bibitem{AvBaLoMeOl12}
P.~P. Avelino, D.~Bazeia, L.~Losano, J.~Menezes, B.~F. Oliveira, Junctions and
  spiral patterns in rock-paper-scissors type models, arXiv:1205.6078 (2012).

\bibitem{AbZa98}
G.~Abramson, D.~H. Zanette, Statistics of extinction and survival in
  lotka-volterra systems, Phys. Rev. E 57 (2010) 4572--4577.

\bibitem{MaMiSnTr11}
J.~Mathiesen, N.~Mitarai, K.~Sneppen, A.~Trusina, Ecosystems with mutually
  exclusive interactions self-organize to a state of high diversity, Phys. Rev.
  Lett. 107 (2011) 188101.

\bibitem{PaZlScCa11}
G.~M. Palamara, V.~Zlatic, A.~Scala, G.~Caldarelli, Population dynamics on
  complex food webs, Adv. Comp. Syst. 14 (2011) 635--647.

\bibitem{FrAb01}
M.~Frean, E.~R. Abraham, Rock=scissors-paper and the survival of the weakest,
  Proc. R. Soc. London B 268 (2001) 1323--1327.

\bibitem{ClTr08}
J.~C. Claussen, A.~Traulsen, Cyclic dominance and biodiversity in well-mixed
  populations, Phys. Rev. Lett. 100 (2008) 058104.

\bibitem{Masuda08}
N.~Masuda, Oscillatory dynamics in evolutionary games are suppressed by
  heterogeneous adaptation rates of players, J. Theor. Biol. 251 (2008)
  181--189.

\bibitem{HeMoTa10}
Q.~He, M.~Mobilia, U.~C. Täuber, Spatial rock-paper-scissors models with
  inhomogeneous reaction rates, Phys. Rev. E 82 (2010) 051909.

\bibitem{VePl10}
S.~Venkat, M.~Pleimling, Mobility and asymmetry effects in one-dimensional
  rock-paper-scissors games, Phys. Rev. E 81 (2010) 021917.

\bibitem{JiZhPeWa11}
L.-L. Jiang, T.~Zhou, M.~Perc, B.-H. Wang, Effects of competition on pattern
  formation in the rock-paper-scissors game, Phys. Rev. E 84 (2011) 021912.

\bibitem{Tainaka93}
K.-I. Tainaka, Paradoxical effect in a three candidate voter model, Phys. Lett.
  A 176 (1993) 303--306.

\bibitem{IfBe03}
M.~Ifti, B.~Bergersen, Survival and extinction in cyclic and neutral
  three-species systems, Eur. Phys. J. E 10 (2003) 241--248.

\bibitem{ReMoFr06}
T.~Reichenbach, M.~Mobilia, E.~Frey, Coexistence versus extinction in the
  stochastic cyclic lotka-volterra model, Phys. Rev. E 74 (2006) 051907.

\bibitem{MaDi99}
J.~Marro, R.~Dickman, Nonequilibrium phase transitions in lattice models,
  Cambridge University Press, Cambridge, 1999.

\bibitem{SzSzIz04}
G.~Szabó, A.~Szolnoki, R.~Izsák, Rock-scissors-paper game on regular
  small-world networks, J. Phys. A: Math. Gen. 37 (2004) 2599--2609.

\bibitem{KiSi94}
V.~Kirk, M.~Silber, A competition between heteroclinic cycles, Nonlinearity 7
  (1994) 1605--1621.

\bibitem{Ga92}
A.~Gaunersdorfer, Time averages for heteroclinic attractors, SIAM J. Appl.
  Math. 52 (1992) 1476--1489.

\bibitem{OvMe10}
O.~Ovaskainen, B.~Meerson, Stochastic models of population extinction, Trends
  Ecol. Evol. 25 (2010) 643--652.

\bibitem{ReMoFr07a}
T.~Reichenbach, M.~Mobilia, E.~Frey, Mobility promotes and jeopardizes
  biodiversity in rock-paper-scissors games, Nature 448 (2007) 1046--1049.

\bibitem{CrReFr09b}
J.~Cremer, T.~Reichenbach, E.~Frey., The edge of neutral evolution in social
  dilemmas, New J. Phys. 11 (2009) 093029.

\bibitem{ScCl10}
M.~Schütt, J.~C. Claussen, Mean extinctin times in cyclic coevolutionary
  rock-paper-scissors dynamics, arXiv:1003.2427 (2010).

\bibitem{MuGa10}
A.~P.~O. Mueller, J.~A.~C. Gallas, How community size affects survival chances
  in cyclic competition games that microorganisms play, Phys. Rev. E 82.

\bibitem{Petraitis79}
P.~S. Petraitis, Competitive networks and measures of intransitivity, Am. Nat.
  114 (1979) 921--925.

\bibitem{LaSc06}
R.~A. Laird, B.~S. Schamp, Competitive intransitivity promotes species
  coexistence, Am. Nat. 168 (2006) 182--193.

\bibitem{RoAl11}
J.~Rojas-Echenique, S.~Allesina, Interaction rules affect species coexistence
  in intransitive networks, Ecology 92 (2011) 1174--1180.

\bibitem{ShMc12}
D.~Shizuka, D.~B. McDonald, A social network perspective on measurements of
  dominance hierarchies, Animal Behaviour 83 (2012) 925--934.

\end{thebibliography}

\end{document}